\documentstyle[epsf,prb,preprint,aps]{revtex}
\tighten
\begin{document}
\draft
\title{Coulomb Blockade Ratchet}
\author{Xiao-hui Wang$^{1}$, Tobias Junno$^{2}$,   
Sven-Bertil Carlsson$^{2}$, \\ 
Claes Thelander$^{2}$ and Lars Samuelson$^{2}$}
\address{$^{1}$ Department of Theoretical Physics, Lund University \\
Helgonav{\"a}gen 5, S-223 62 Lund, Sweden\\
$^{2}$Solid State Physics/The Nanometer Consortium, \\
Lund University, P.~O.~Box 118,\\
S-221 00 Lund, Sweden}
\date{\today}
\maketitle
\begin{abstract}
We investigate the transport properties of a new class of ratchets.
The device is constructed by applying an ac voltage to the metallic single
electron tunneling transistor, and a net transport current is induced by the 
time-dependent bias-voltage, although the voltage value is on the average 
zero.  The mechanism underlying this phenomenon is the Coulomb blockade of 
the single electron tunneling.  The directions and the values of the induced 
net currents can be well controlled by the gate-voltage.  A net transport 
current has also been observed even in the absence of the external 
bias-voltages, which is attributed to the noise in the circuit.

\end{abstract}


The principle of mechanical ratchets has been analyzed in the course of the
development of modern physics.\cite{smo14,fey66}  Recently, the idea of
Feynman's thermal ratchet has been generalized\cite{but87,han97,mil95} to
account for the macroscopic motion of particles in a unbiased asymmetric
periodic potential such as material transport in biological
systems.\cite{ish91} A quantum rectifier using the superposition of a 
sinusoidal oscillation and its second harmonic with a phase shift as
bias-voltage applied to a periodic symmetric potential has been predicted to
be able to create a net current flow.\cite{han97}  The prediction has been
confirmed by experiments using a 2D-array of triangular-shaped
anti-dots.\cite{lor98} A ballistic rectifier using a single anti-dot as an
asymmetric artificial scatterer in a semiconductor microjunction has been
demonstrated experimentally to guide carriers in a predetermined spatial
direction, thus behaving like a four-diode-bridge rectifier.\cite{song98}
A geometric quantum ratchet has also been realized by applying an ac
voltage bias to a triangular-shaped quantum dot to create a net 
current.\cite{link98} In these two cases, there are no periodic potentials
in the system.  Instead, the charge carriers move randomly in an
asymmetric structure in the direction of the transport current, and
can drift out to create a current.
In this letter, we present our investigation of a novel class of
ratchets.  In our device, we do not need external electrostatic
potentials to confine the charge carriers, namely electrons, neither do we 
need specially prepared geometrical confinement.  The central part
of our ratchet is formed by a metallic grain of arbitrary shape, which
is brought within a distance of a couple of \AA ngstrom from two
metallic leads.\cite{tob98,sve99}
Electrons can tunnel between the leads and the grain.  If the grain is
sufficiently small, and coupled properly to voltage sources, such a
device is actually the single electron tunneling transistor (SET),
where the central grain is called the island, which is coupled to the
gate voltage $V_g$ via the gate capacitance $C_g$, and the leads are
directly coupled to the transport voltages.

The transport current as a function of dc bias- and gate-voltage has been
extensively investigated both theoretically and experimentally in
recent years.\cite{tob98,sve99,ave91,fd87} It becomes clear that in
such a device with sufficiently large tunnel resistances at
sufficiently low temperatures, tunneling of even a single electron is
not allowed for vanishing gate voltage as long as the dc bias-voltage
is smaller than the threshold value $V_c=e/2(C_1+C_2+C_g)$, where
$C_1$ and $C_2$ are the capacitances of the tunnel junctions.  As a
result, there is a finite bias-voltage, but no current flows.  This
phenomenon is known as the Coulomb blockade, which can be lifted out
by increasing the transport voltage or tuning the gate voltage.  
Distinct Coulomb staircases in the $I-V$ curves have been observed in our 
device when the parameters of the two tunnel junctions are made different 
from each other.\cite{tob98,sve99} For the SET used in our present 
investigation, the $I-V$ curve is shown in Fig.~1  for  $V_g=0$ and 
$T=4.2 K$.  The corresponding parameters of the device are found to be 
$C_1=7.6$ aF, $R_1=1.0$ M$\Omega$, $C_2=7.6$ aF, $R_2=105.0 $ M$\Omega$, 
$C_g=2.0 $ aF, and $E_c=4.7$ meV. 

If the circuit is biased by an ac voltage, the average bias voltage is zero,
the average transport current would be, {\it prima facie},
also zero just as in the case of vanishing transport voltage.  This is
certainly true for classical tunnel junctions, where the capacitances of the
junctions are large, and thus the Coulomb charging energies are
negligible compared to the thermal energy even at fairly low temperatures
so that the I-V characteristics are Ohmic.
For our device with the charging energy of $4.7 meV$, however, the I-V
characteristics at liquid helium temperature are strongly nonlinear due to
the Coulomb blockade,  and the relation $I(-V)=-I(V)$ is in general not
satisfied. Therefore the net current is nonvanishing, even if the average
transport voltage is zero. To get a pronounced net current induced by the ac 
bias-voltage, the two tunnel junctions should have different parameters 
so that the I-V curves are strongly asymmetric with respect to 
the applied voltages. In the experiment reported here, we choose 
$R_1 \ll R_2$, and indeed we have observed clearly transport currents induced 
by ac bias-voltages, which have well-defined lineshapes extended periodically
for a wide range of the gate-voltage.  Moreover, we can control both the
direction and the magnitude of the net currents by tuning the gate-voltage.
The experimental data are found to agree very well with the theoretical
calculations based on the constant charging energy model.\cite{ave91}
Since the only physical reason that we obtain the net current by applying
ac bias-voltage is the Coulomb blockade of the single electron tunneling,
we call our device a {\it Coulomb blockade ratchet}.

The rectifier effect in a lateral 2D quantum dot SET made in a semiconductor 
nanostructure has been observed by Weis {\it et al}.\cite{wei95} 
However, in a semiconductor SET\cite{mf95} the transport cannot be simply
attributed to the Coulomb blockade, but depends on the details of the
device such as material, external confinement, and coupling of the
energy levels in the quantum dot to the leads.  On the contrary, our device 
is merely characterized by junction resistances, junction capacitances, and 
gate capacitance, and the induced net current is perfectly periodic in the
gate-voltage, and determined explicitly by the above-mentioned parameters 
of the device.  In addition to the case of the ac biased circuit, we have
also studied the transport properties of the SET driven by noise
sources, which lead to detectable net currents, even in the absence of
any applied ac voltages.
The transport theory of the SET in the parameter range of our devices
biased by dc voltages is well-established.\cite{ave91}  The dc current of
the SET can be calculated via
\begin{eqnarray} \label{iv}
& & I(V)=e\sum_{n=-\infty}^{\infty} \sigma(n, V)
[\Gamma_1^+(n, V)-\Gamma_1^-(n, V)]
\nonumber \\
& &=-e\sum_{n=-\infty}^{\infty} \sigma(n, V)
[\Gamma_2^+(n, V)-\Gamma_2^-(n, V)] \, .
\end{eqnarray}
Here the tunneling rates of an electron tunneling from (-) or onto (+) the
central island with n excess electrons via the first (1) or the second (2)
junction are labeled by $\Gamma_{1,2}^{\pm}(n, V)$, and the probability of
$n$ excess electrons on the island is denoted by $\sigma(n, V)$.
Since the resistances of the tunnel junctions in our experiments are typically
of the order of 100 $M\Omega$, and the capacitances are of the order of 10 
$aF$, the RC time is thus of the order of $10^{-9} s$, which is much larger 
than the tunneling time.  The latter is of the order of $10^{-15} s$ for 
metallic tunnel junctions.\cite{but82}  In the experiments, we have varied the
frequencies of the ac voltages $\omega/2\pi$ between $100 Hz$ and  $10 MHz$.
In this frequency range the period of the bias signal ${\cal T}=2\pi/\omega$
is much larger than the intrinsic characteristic time-scales of the SET, such
as electron tunneling time or capacitance charging time, so that the SET can
follow the variation of the ac signals adiabatically.  Hence the
induced net current is given by the mean value of the transport voltage, 
$I_{\rm net}=\langle I[V(\varphi)] \rangle$,
where $I[V(\varphi)]$ is given by Eq.~\ref{iv} with
$V(\varphi)=V_{\rm am} \cos (\varphi)$ and $\varphi =\omega t$ .
Since $I[V(\varphi)]$ is a periodic function of time, the mean value of it can 
be calculated either within a time interval much larger that the period 
${\cal T}$, or within a period of the applied ac voltage. 
Apparently, the net current expressed in the above
formula has no explicit frequency-dependence, which is confirmed by our
experiments for various frequencies between $100 Hz$ and $10 MHz$.

The fabrication and operation of our SET has been published
elsewhere,\cite{tob98,sve99} and will not be repeated here.  The only
difference is that the island and the leads in the present device are made of 
palladium instead of gold. The tunnel gaps were
deliberately tuned to be strongly asymmetric in resistance in order to
enhance the ratchet effect.
To give a comprehensive description of the device, we have
investigated the net current as a function of the gate voltage for various
amplitudes of the ac bias voltages.  For large amplitudes of the 
bias-voltages, we found that by decreasing the amplitude of the
ac bias-voltages, the amplitudes of the induced currents 
get smaller rapidly, and the lineshapes shrink.  Typical curves
are shown in Fig.~2 (dots) for large amplitude of the applied ac voltage 
where the lineshapes are sine-like, and in Fig.~3 (dots) for intermediate 
amplitude of the applied ac voltage where the variation of the net current is 
relatively slow in the Coulomb blockade regime.  In the figures the 
corresponding theoretical curves of the net current induced by the applied 
bias-voltage as a function of the gate-voltage 
calculated according to Eq.~\ref{iv} are also shown (dashed lines).
By further decreasing amplitude of the applied ac voltage the induced
net current is expected to decrease and finally vanish for zero
amplitude. 
However, in the experiments, we observed a fairly pronounced transport current
as a periodic function of the gate-voltage, even when our applied ac
bias-voltage was vanishing, as shown by the dots in Fig.~4. 
We explain this by considering that the input terminal of the device 
picks up the external noise that acts as a background, time-dependent
bias-voltage with zero average values. In our experimental setup this is 
reasonable because there are coaxial cables several meters long  
between the instruments/filtering at room temperature, and the sample.
The presence of noise sets the lower limit for the amplitude of the 
induced current.

By measuring the input noise to the SET in the frequency range of $10 Hz$ to 
$100 kHz$, we found that the dominating noise source has a broad spectrum   
with almost constant intensity smaller than $10^{-4} mV/\sqrt{Hz}$.  
However this broad band noise is unlikely to induce the measured net current, 
because the induced net current depends strongly on the amplitude of the ac 
bias-voltage. For the parameters of our device, the net current induced by 
the applied ac bias-voltage becomes smaller in the order of magnitude if the 
amplitude of the ac bias-voltage decreases from a few $mV$ to a value 
smaller than $1 mV$.  This is at least one order of magnitude 
larger than the rms value of the broad band noise voltage in our circuit.  
Even if the rms value of the broad band noise is larger than $1 mV$, the 
amplitude  of the signal in each small frequency interval is still very 
small, while the bandwidth of the noise spectra remains large.  As a 
consequence, the contribution of the broad band noise to the induced net 
current is negligible. To consolidate this argument, we have deliberately 
applied a broad band noise source with the rms value of input voltage as 
large as $8.5 mV$ to the device.  As shown by the crosses in Fig.~4, 
this additional noise source has indeed little effect on both the amplitude 
and the lineshape of the net current as compared to the one of the unbiased 
device.

We attribute the net current in the unbiased device to the noise source 
with large amplitudes in the high frequency range.  
The measured data can be fitted reasonably
well by the standard sequential tunneling theory\cite{ave91} if one chooses
the amplitude of the fitting ac bias-voltage to be $2.3 mV$, as shown by the
dashed line in Fig.~4.  Furthermore, we have calculated the induced net
current as a function of the gate-voltage in the presence of both the applied 
ac voltage and the circuit noise voltage. 
The circuit noise is modeled by the ac voltage with the amplitude obtained 
from Fig.~4, and with the frequency much higher than that of the applied 
voltage.  Then the current for a given point of the applied voltage is 
calculated as the mean value with respect to the circuit noise in a time 
interval much smaller than the period of the applied voltage, but much larger 
than the period of the modeled ac voltage of the circuit noise, i.e.\ 
$\tilde{I}[V(\varphi)]=\langle I[V_{\rm am} \cos (\varphi)
+\tilde{V}_{\rm am} \cos (\tilde{\varphi})] \rangle$,
and the observed current is thereafter calculated as the mean value with 
respect to the applied ac signal, 
$\tilde{I}_{\rm net}=\langle \tilde{I}[V(\varphi)] \rangle$.
The results are shown by the solid lines in Fig.~2 and
Fig.~3.  For a fairly large amplitude of the ac
bias-voltage, the influence of the circuit noise is weak, as shown 
in Fig.~2, while for an intermediate amplitude of the ac bias-voltage, the 
influence is significant.  As shown in Fig.~3, the result of the
ac plus noise bias-voltage agrees very well with the experimental data for
the whole curve, while the one of the pure ac bias-voltage gives the
correct amplitude and the lineshape near to the resonance at $C_g V_g=e/2$
module e, yet a somewhat smaller value near to the Coulomb blockade at
$C_g V_g=0$ module e.  From the above analysis, it seems that the measured 
unbiased net current is very likely to be induced by the circuit noise with 
large amplitudes in the high frequency range.

In summary, we have investigated the Coulomb blockade ratchet
using the metallic single electron tunneling transistor.  We have observed 
fairly long periods of the net currents induced by the ac bias-voltages, 
which agree very well with the sequential tunneling theory.   In the
absence of applied voltages, we also observed pronounced, periodic net 
currents, which are probably induced by the circuit noise.  We have hence 
shown how the background noise is, in a ratchet-like fashion, transformed 
by the single electron device into a net current of electrons.    

The authors would like to thank H.~Linke for valuable comments on our 
manuscript, and A.~L{\"o}fgren, P.~Omling, and H.~Xu for stimulating 
discussions.

 
\begin{figure}
\caption{I-V curve of the SET at $4.2 K$ and $V_g=0$ with the device 
parameters $C_1=7.6 aF$, $R_1=1.0 M\Omega$, 
$C_2=7.6 aF$, $R_2=105.0 M\Omega$ and $C_g=2.0 aF$.}
\label{fig1}
\end{figure}
 
\begin{figure}
\caption{Induced current as a function of the gate voltage for a large 
applied ac voltage with the rms value of $6 mV$.  The dots 
are the experimental data, while the solid and dashed lines correspond to 
the theoretical results with or without the circuit noise contributions, 
respectively.}
\label{fig2}
\end{figure}

\begin{figure}
\caption{Induced current as a function of the gate voltage for an 
intermediate applied ac voltage with the rms value of $3 mV$.  
The dots are the experimental data, while the solid and dashed lines 
correspond to the theoretical results with or without the circuit noise 
contributions, respectively.  Note the dashed line is less smooth near the 
Coulomb blockade, which is smeared out by the circuit noise to become  
the solid line.  The latter agree very well with the experimental data.} 
\label{fig3}
\end{figure}

\begin{figure}
\caption{Induced current as a function of the gate voltage in the absence of 
the applied ac voltage.  The dots are the experimental data 
without additional broad band noise, the dashed curve is the theoretical 
fitting with an amplitude of the ac bias-voltage of $2.3 mV$.  The crosses 
are the experimental data when the device is biased by a broad band noise 
voltage with the rms value of $8.5 mV$.}
\label{fig4}
\end{figure}


\begin{references}

\bibitem{smo14}
M.~von Smoluchowski, in {\it Vortrage \"uber die kinetische Theorie der
Materie und der Elektrizit\"at} edited by M.~Planck (Teubner und Leipzig,
Berlin, 1914).
\bibitem{fey66}
R.~P.~Feynman, R.~B.~Leighton and M.~Sands, {\it The Feynman LECTURES ON
PHYSICS}, Chapter 46, (Addison-Wesley, Reading, MA, 1966).
\bibitem{but87} 
M.~B{\"u}ttiker, Z.~Phys.~ {\bf 68}, 161 (1987);
R.~Landauer, J.~Stat.~Phys.~{\bf 53}, 233 (1988).
\bibitem{han97}
P.~Reiman, M.~Grifoni and P.~H\"anggi, Phys.~Rev.~Lett.
{\bf 79}, 10 (1997); I.~Goychuk and P.~H\"anggi, Europhys.~Lett.~{\bf 43},
503 (1998).
\bibitem{mil95}
M.~M.~Millonas,  Phys.~Rev.~Lett.~{\bf 74}, 10 (1995).
\bibitem{ish91}
D.~A.~Astumian, Science {\bf 276}, 917 (1997); Y.~Okada and N.~Hirokawa, 
{\it ibid} {\bf 283}, 1152 (1999).
\bibitem{lor98}
A.~Lorke, S.~Wimmer, B.~Jager, J.~P.~Kotthaus, W.~Wegscheider and 
M.~Bichler, Physica B {\bf 249-251} 312 (1998).
\bibitem{song98}
A.~M.~Song, A.~Lorke, A.~Kriele and J.~P.~Kotthaus, Phys.~Rev.~Lett.
{\bf 80}, 3831 (1998).
\bibitem{link98}
H.~Linke, W.~Sheng, A.~L\"ofgren, H.~Xu, P.~Omling and P.~E.~Lindelof, 
Europhys.~Lett.~{\bf 44}, 341 (1998); H.~Linke, T.~E.~Humphrey, 
A.~L\"ofgren, A.~O.~Sushkov, R.~Newbury, R.~P.~Taylor and P.~Omling, 
to be published. 
\bibitem{tob98}
T.~Junno, S.-B.~Carlsson, H.~Xu, L.~Montelius, and L.~Samuelson,
Appl.\ Phys.\ Lett.\ {\bf 72}, 548 (1998).
\bibitem{sve99}
S.-B.~Carlsson, T.~Junno, L.~Montelius, and L.~Samuelson,
Appl.\ Phys.\ Lett.\ {\bf 75}, 1461 (1999).
\bibitem{ave91}
D.\ V.\ Averin and K.\ K.\ Likharev in {\it Mesoscopic Phenomena in Solids},
edited by B.\ L.\  Altshuler, P.\ A.\ Lee, and R.\ A.\ Webb (Elsevier,
Amsterdam, 1991).
\bibitem{fd87}
T.\ A.\  Fulton G.\ J.\ Dolan, Phys.\ Rev.\ Lett.\ {\bf
59},  109 (1987).
\bibitem{wei95}
J.~Weis, R.~J.~Haug, K.~von Klitzing and K.~Ploog, Semicon.\ Sci.\ Technol.\
{\bf 10}, 877 (1995).
\bibitem{mf95}
U.~Meirav and E.~B.~Foxman, Semicon.\ Sci.\ Technol.\
{\bf 11}, 255 (1995).
\bibitem{but82}
M.~B{\"u}ttiker and R.~Landauer, Phys.~Rev.~Lett.~{\bf 49}, 1739 (1982).

\end{references}
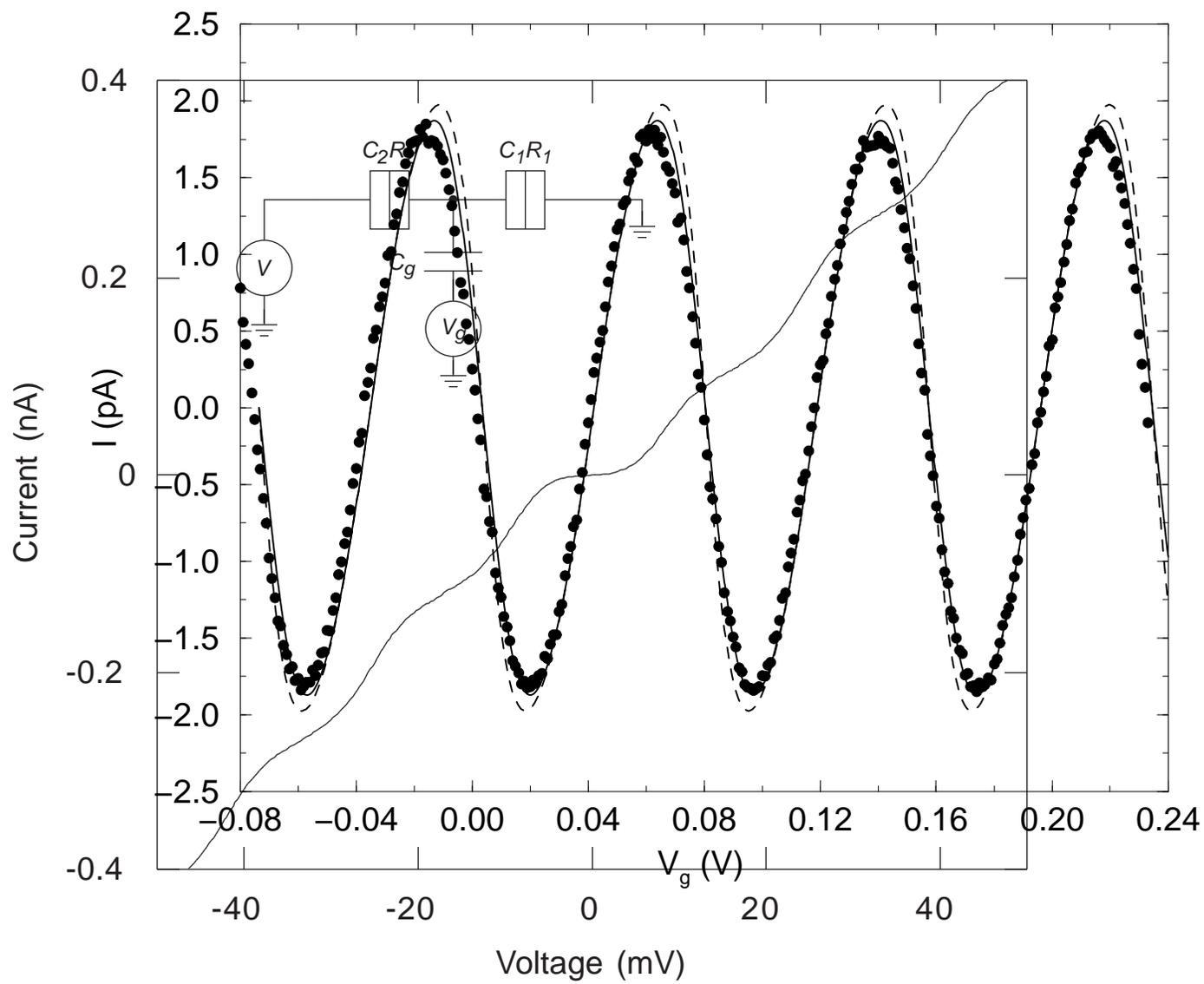
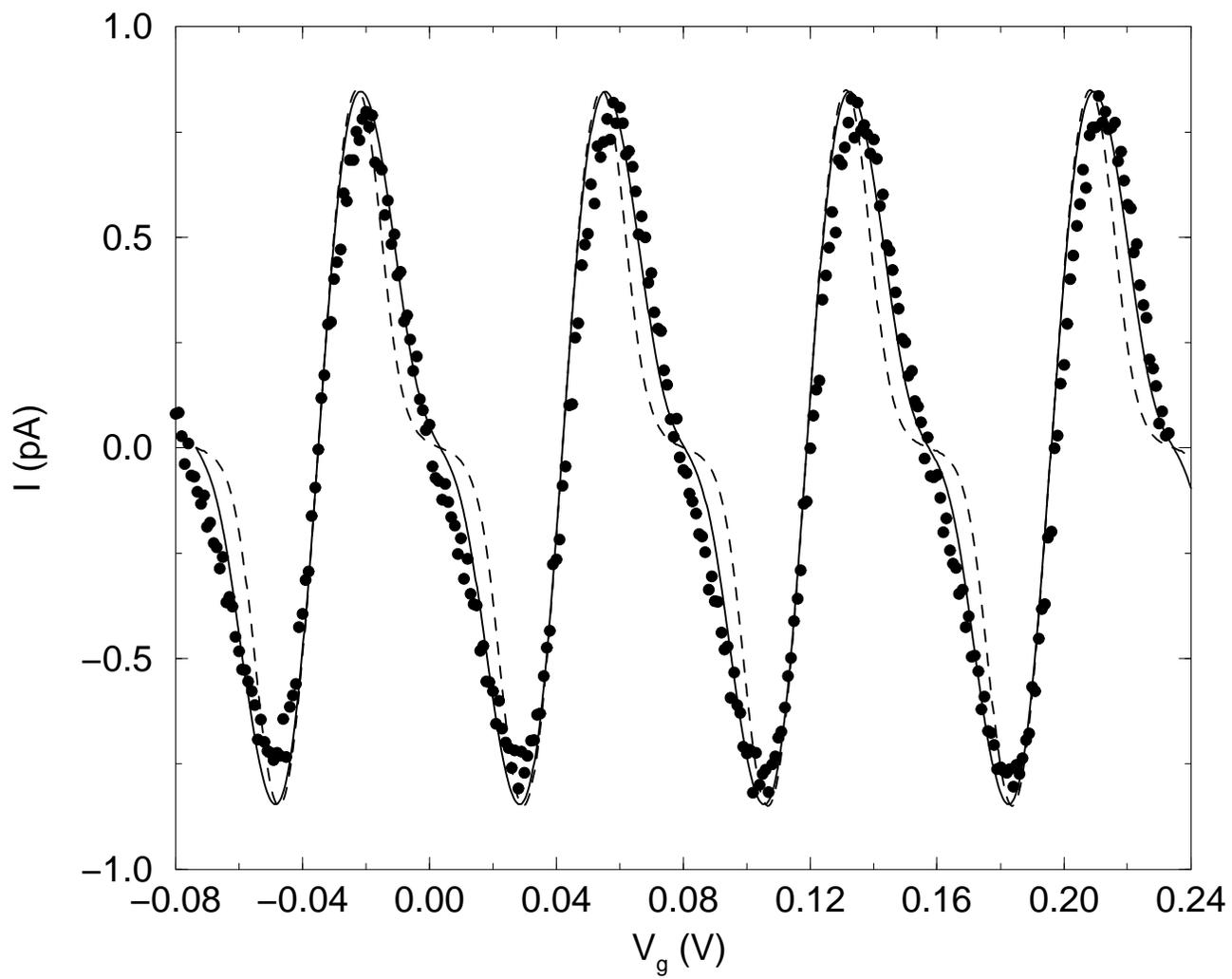
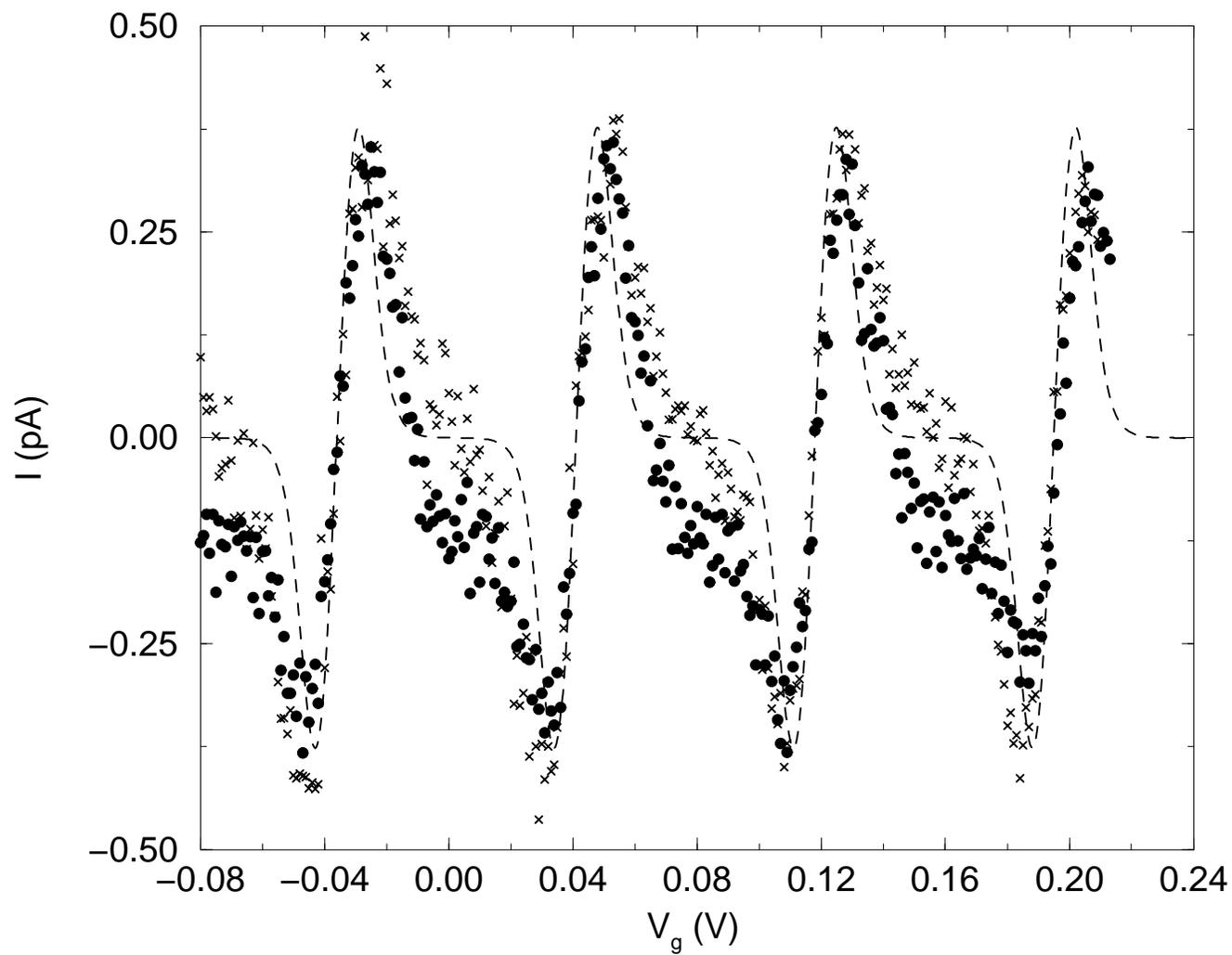
\end{document}